# Conversational Factor Information Retrieval Model (ConFIRM)

**Stephen Choi, William Gazeley, Siu Ho Wong, Tingting Li**
IRAI Labs, LORA Research
`{stepchoi, william}@irai.co {adrian, tingting}@asklora.ai`

## Abstract

This paper introduces the Conversational Factor Information Retrieval Method (ConFIRM), a novel approach to fine-tuning large language models (LLMs) for domain-specific retrieval tasks. ConFIRM leverages the Five-Factor Model of personality to generate synthetic datasets that accurately reflect target population characteristics, addressing data scarcity in specialized domains. We demonstrate ConFIRM's effectiveness through a case study in the finance sector, fine-tuning a Llama-2-7b model using personality-aligned data from the PolyU-Asklora Fintech Adoption Index. The resulting model achieved 91% accuracy in classifying financial queries, with an average inference time of 0.61 seconds on an NVIDIA A100 GPU. ConFIRM shows promise for creating more accurate and personalized AI-driven information retrieval systems across various domains, potentially mitigating issues of hallucinations and outdated information in LLMs deployed in mission-critical environments.

## 1 Introduction

The rapid advancement of large language models (LLMs) has been remarkable, with significant breakthroughs in scaling laws (Kaplan et al., 2020) and subsequent performance gains (Zhao et al., 2023). However, deploying LLMs in mission-critical domains such as healthcare, law, and finance poses unique challenges that go beyond general AI optimization and alignment. One major concern is the tendency of LLMs to produce convincing yet inaccurate responses, known as "hallucinations," particularly in specialized areas where data is scarce or nuanced.

Additionally, the exponential growth of online data, coupled with the significant resources required for data annotation and model training, hinders the ability of LLMs to remain current. Recent advancements aim to address these shortcomings. Retrieval-Augmented Generation (RAG) (Lewis et al., 2020) integrates information retrieval (IR) methods to mitigate the limitations of static knowledge. Chain of Thought prompting (Wei et al., 2022) has spurred the development of task-based workflows and agentic pipelines. These specialized agents are typically trained efficiently on a small subset of the model's parameters that significantly impact the task at hand, utilizing parameter-efficient fine-tuning (PEFT) methodologies (Lialin et al., 2023), such as Low-Rank Adaptation (LoRA) (Hu et al., 2021) and Reinforcement Learning from Human Feedback (RLHF) (Ouyang et al., 2022).

However, these training methods require substantial-sized datasets. For instance, the RLHF methods outlined by Ouyang et al. (2022) involved generating a 31,000-sample preference dataset for the reward model, underscoring the need for large-scale data in fine-tuning processes. These methods also perform optimally when the training dataset accurately represents the target population's characteristics. The Conversational Factor Information Retrieval Method (ConFIRM) is designed to assess these characteristics and efficiently synthesize questions in scale for effective fine-tuning. We outline this method in Section 2.

In Section 3, we apply ConFIRM to a real-world example using survey data from Hong Kong Polytechnic University's PolyU-Asklora Fintech Adoption Index[1] to address a financial scenario. While we do not include the full information retrieval instructions due to scope limitations, we focus on classification to highlight the practical business case. This involves the "retrieval" components of RAG, rather than additional engineering required to handle units, currency, and other numerical issues. To aid reproduction, we post our model code at: https://github.com/WilliamGazeley/ConFIRM

---

[1] https://www.polyu.edu.hk/en/media/media-releases/2022/1102_polyus-school-of-accounting-and-finance-launches--its-first-fintech-adoption-index-survey

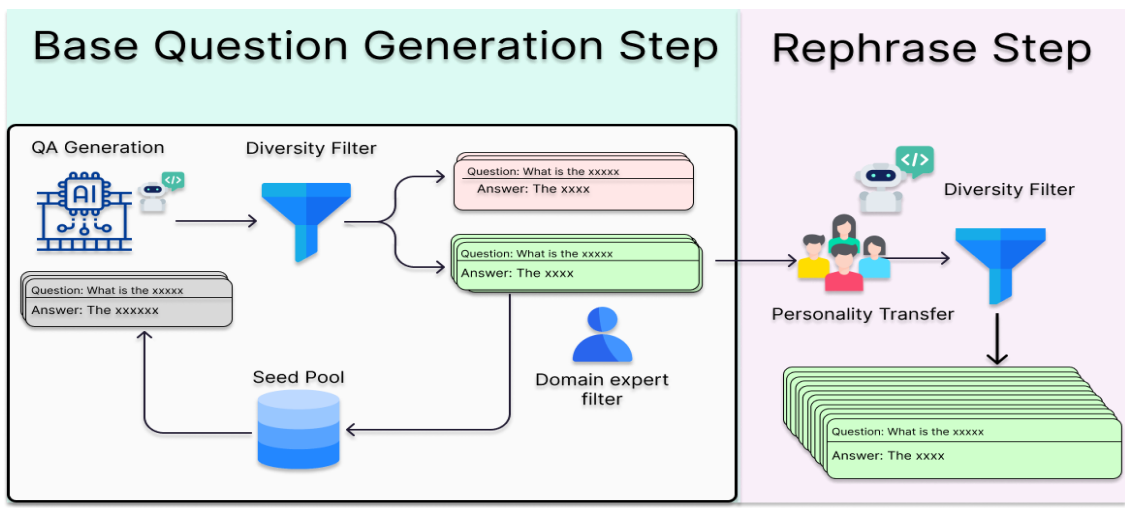

Figure 1: A schematic illustration of question-answer generation process

## 2 Theoretical Framework

To create a synthetic dataset that accurately reflects personality traits within a population, we utilize the Five-Factor Model (FFM) of personality (McCrae and Costa, 1999). The FFM quantifies personality across five dimensions: Openness, Conscientiousness, Extraversion, Agreeableness, and Neuroticism (OCEAN). This model is supported by extensive research, including behavioral genetics studies confirming its structural integrity (Yamagata et al., 2006), longitudinal studies demonstrating its stability over time (Roberts and DelVecchio, 2000), and cross-cultural research affirming its universal applicability (John et al., 2008). Additionally, significant associations have been established between OCEAN traits and key financial outcomes (Exley et al., 2021), as well as their utility in financial planning (Campbell et al., 2023).

We utilize the 50-question set from International Personality Item Pool (IPIP)[2]. Instead of employing the varimax factor loading scoring method (Goldberg, 1992), we track the highest scoring domain (factor) for each participant. Subsequently, we generate question-answer pairs by leveraging Large Language Models (LLMs). To craft natural-sounding questions relevant to our task, we employ the iterative instruction generation approach of SELF-INSTRUCT (Wang et al., 2023), aiming to synthesize authentic user interactions. Chat GPT-3.5 (Brown et al., 2020) aids in generating conversational questions.

Through an iterative process of manual filtering and refinement, we select high-quality question-answer pairs. To further emulate a conversational tone, we apply the Text2Text generation method (Ramirez et al., 2023). A few-shot prompt, engineered based on OCEAN FFM max-factor scores, integrates population-aligned factor loadings into the questions. Pseudo-references, derived from meaning representations, are used with an LLM (GPT-3.5) for few-shot in-context learning to generate questions in an extraverted style. This approach efficiently generates a large, diverse training set required for robust fine-tuning task-oriented agents. The final set can be filtered using ROUGE-L scores (Lin and Och, 2004) both vertically (across questions) and horizontally (within pairs) to enhance diversity and reduce redundancy. The question-answer generation process is outlined in Figure 1.

## 3 Case Study

In this section, we present a practical implementation of the ConFIRM framework through a case study in the finance sector. The objective is to fine-tune a Retrieval-Augmented Generation (RAG) model to effectively retrieve information from an internal database based on questions from a target population. Our goal is to generate a large dataset of questions-answer pairs to fine-tune an agent to accurately label the data category necessary to answer each question. For this case study, we focus primarily on stocks, but this approach can easily be extended to include additional or different data categories.

---

[2] https://ipip.ori.org/new_ipip-50-item-scale.htm

| OCEAN Factor | % max factor |
|---|---|
| Openness | 10% |
| Conscientiousness | 36% |
| Extraversion | 14% |
| Agreeableness | 18% |
| Neuroticism | 22% |

Table 1: Sample OCEAN Max Factors

### 3.1 Retrieval Database

The internal retrieval database structure is modeled after **Refinitiv Datastream**, a leading and widely used provider of financial data. Refintiv offers a comprehensive range of global financial data from various sources, including stock exchanges, central banks, and other financial institutions. The platform also provides information on the data category, data types, and descriptions of each data field. Using this foundation, we structure the following data categories for a hypothetical stock investment firm. (For detailed information on the actual fields, please refer to Appendix A.)

**Stock data:** This category is a table that contains data about individual stocks, including price data, identifiers, descriptions, and data available in company filings. For the scope of this paper, we limit the number of data fields to the top 40 ranked by popularity within the category marked as "Equities" by Refinitiv.

**Market data:** This category holds data about the overall stock market, including benchmark indices such as S&P 500, and sector and style benchmarks. We limit the number of data fields to the top 15 ranked by popularity within the "Equity Indices" category.

**Economic data:** This category holds data and statistics about the US economy as a whole, such as GDP, unemployment rate and consumer price index as well as leading indicators. We limit the number of data fields to the top 11 ranked by popularity within the "Economics" category.

**News data:** This category is a data warehouse where all news scraped from Reuters (one of the largest news agencies) is stored.

**External data:** This category represents the external knowledge base where all data that is out-of-scope of the four internal categories. In terms of label accuracy, this category represents the "true negatives". A typical example of this data source is social media or internet sites.

| Hyperparameter | LoRA |
|---|---|
| learning rate | 3E-04 |
| batch size | 8 |
| epochs | 50 |
| max length | 128 |
| r | 4 |
| dropout | 1.00E-03 |
| alpha | 64 |

Table 2: LoRA hyperparameters

### 3.2 Aligned QA generation

For the case study, we leveraged the HK PolyU-Asklora Fintech Adoption Index survey to measure the OCEAN factors of potential users in Hong Kong. Although the survey included 1000 participants, only a subset of 50 was presented with the IPIP questions, resulting in the percentages shown in Table 1.

Using the methodology described in Section 2, we generated question-answer pairs, which were then expanded to match the personality factor ratios of the sample (Table 1). This process resulted in 3300 samples reflecting the target population. We allocated 3000 samples for training and 300 for validation (10% of the training set). To avoid data leakage, we generated an additional 1000 samples using the same process for testing purposes. Refer to Appendix B for examples of the generated samples.

### 3.3 Fine-Tuning Agent

With the training and test sets prepared, we focus on parameter-efficient fine-tuning of a manageable pre-trained LLM, Llama-2-7b (Touvron et al., 2023). This arrangement allows full control of the model location and network. The goal is to convert a general-purpose base model (Llama-2-7b) to a specialized classifier. We retain the weights of the pre-trained model and apply Low-Rank Adaptation (LoRA) (Hu et al., 2021) for fine-tuning. We utilize the validation set to optimize the hyperparameters (Table 2).

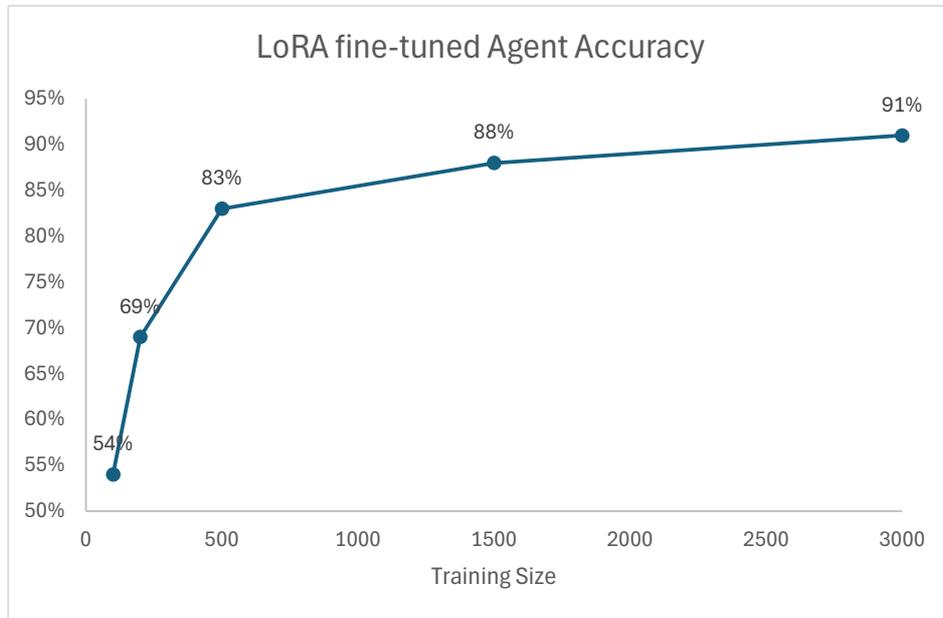

Figure 2 Data Efficiency of Agent

### 3.4 Results

Accuracy is evaluated on two factors: accuracy of internal/external data category labeling and precision of data category and field labels. Instances where false negatives occurred in external labels are marked as failed classification, underscoring potential regulatory concerns. For the purposes of the case study, a classification is determined as successful if the fine-tuned agent returned at least a superset of the correct data category or categories within the answer set.

The need for a large sample size is illustrated in Figure 2. The LoRA fine-tuned agent only surpassed 80% accuracy after 500 training samples and surpassed 91% accuracy with 3000 training samples. The average runtime for classification (inference) on each test question for the LoRA-tuned model on a single NVIDIA A100 GPU is 0.61 seconds.

### 4 Conclusion

In this paper, we introduced Conversational Factor Information Retrieval Method (ConFIRM), a novel approach to synthetically generating training samples to fine-tune large language models for domain-specific retrieval tasks. By leveraging the Five-Factor Model of personality and advanced language model techniques, ConFIRM addresses the critical challenge of creating representative synthetic datasets for specialized domains.

Our case study in the finance sector demonstrated the efficiency of ConFIRM to generate training samples for fine-tuning a Llama-2-7b model for accurate information retrieval and classification. The results show that with sufficient training data generated using our method, we achieved over 91% accuracy in classifying financial queries, highlighting the potential of this approach for real-world applications. While our focus was on the finance sector, the ConFIRM framework has potential applications across various domains where personalized information retrieval is crucial, such as healthcare, legal services, and customer support.

Future work could examine the scalability of this approach to even larger language models, expand on the FFM integration, and explore how ConFIRM can be adopted to other human preference optimization techniques such as Direct Preference Optimization (DPO) (Rafailov et al., 2023).

# A  Appendix – Data Categories and Labels

```
economic_indicators.10Y_ust_yield
economic_indicators.conf_board_leading_indicators
economic_indicators.consumer_confidence
economic_indicators.cpi
economic_indicators.cpi
economic_indicators.fed_funds_rate
economic_indicators.gdp
economic_indicators.id
economic_indicators.industrial_production
economic_indicators.ism_mfg
economic_indicators.ism_non_mfg
economic_indicators.personal_consumption_expenditure
economic_indicators.unemployment_rate
external_data_source
market_data.Communications_services
market_data.Consumer_Disc
market_data.Consumer_Staples
market_data.Energy
market_data.Financials
market_data.Health_Care
market_data.id
market_data.Industrials
market_data.IT
market_data.Materials
market_data.Real_Estate
market_data.S&P500
market_data.Utilities
news_data_source
stock_data.assets_short_term
stock_data.assets_total
stock_data.bps_book_value_per_share
stock_data.cogs
stock_data.common_shares_outstanding
stock_data.company_name
stock_data.date_ex_dividend
stock_data.dividend_yield
stock_data.ebitda
stock_data.employees
stock_data.enterprise_value
stock_data.eps_forward_12_month
stock_data.eps_trailing_12_month
stock_data.general_industry_classification
stock_data.id
stock_data.interest_income
stock_data.interest_income_not_banks
stock_data.inventories_days_held
stock_data.liabilities_short_term
stock_data.liabilities_total
stock_data.long_term_debt
stock_data.market_capitalization
stock_data.operating_expense
stock_data.operating_expense_sg&a
stock_data.operating_income
stock_data.price_close
stock_data.price_intraday_high
stock_data.price_intraday_low
stock_data.ratio_pe
stock_data.research_and_development
stock_data.sales_or_revenues
stock_data.shareholders_equity
stock_data.ticker_symbol
stock_data.total_return
```

## B  Appendix – QA Generation Examples

Base question generation from data categories and labels

> **Few-Shot Prompt**
>
> You are a everyday investor who is member of the general public not a professional investor. Generate a question that can be answered by and only by the following information source or database field. Use informal conversational tone. You MUST NOT include names of the available.
>
> Information:
> Chevron
> stock_data.research_and_development - RESEARCH AND DEVELOPMENT EXPENSE represents all direct and indirect costs related to the creation and development of new processes, techniques, applications and products with commercial possibilities.
> Question: Does Chevron allocate funds for research and development?
>
> Information:
> news_data - The news data is a data warehouse where we store all news scraped from Reuters. Each news item has keywords and date.
> Question: Has there been any recent news about Nvidia?
>
> Information:
> stock_data.ticker_symbol - TICKER SYMBOL represents a symbol used to identify the company on the stock exchanges where it is listed.
> Question: What is the ticker symbol for Apple Inc.?
>
> Information:
> news_data - The news data is a data warehouse where we store all news scraped from Reuters. Each news item has keywords and date.
> Question:
>
> **Model Output**
>
> Is there any news about the recent merger between Disney and Fox?

Questions transformed to incorporate OCEAN personality

**Few-Shot Prompt**

Here is some question: {nameVariable 20-25 decent familyFriendly}. Here is a rewrite of the question which is {UNCONSCIENTIOUSNESS} : {Oh gosh oh God I don't know. nameVariable is family friendly with a decent rating, also it has a price range of £20-25, I mean.}.

Here is some question: {nameVariable coffee shop Indian cheap average familyFriendly}. Here is a rewrite of the question which is {UNCONSCIENTIOUSNESS} : {Oh God err... I don't know. It's cheap and an Indian place, also nameVariable is a coffee shop, and it has an average rating. obviously, nameVariable isn't family friendly.}.

Here is some question: {nameVariable restaurant Japanese mediocre}. Here is a rewrite of the question which is {UNCONSCIENTIOUSNESS} : {Oh gosh I don't know buddy. nameVariable is a Japanese place with a damn mediocre rating, also it is a restaurant.}.

Here is some question: {nameVariable Italian cheap average riverside familyFriendly nearVariable}. Here is a rewrite of the question which is {UNCONSCIENTIOUSNESS} : {I don't know. nameVariable has an average rating, also nameVariable is damn cheap in riverside, also it is an Italian place kind of near nearVariable, and it is darn family friendly.}.

Here is some question: {nameVariable fast food low familyFriendly}. Here is a rewrite of the question which is {UNCONSCIENTIOUSNESS} : {I am not sure. nameVariable has like, a low rating, also it is a fast food place, and nameVariable isn't damn kid friendly friend.}.

Here is some question: {nameVariable coffee shop Japanese 20-25 high nearVariable}. Here is a rewrite of the question which is {UNCONSCIENTIOUSNESS} : {Oh God err... I don't know. I mean, nameVariable has a price range of £20-25 with a high rating, and it is a coffee shop, also nameVariable is a Japanese place near nearVariable.}.

Here is some question: {nameVariable Chinese high average nearVariable}. Here is a rewrite of the question which is {UNCONSCIENTIOUSNESS} : {Yeah, oh God I don't know. Err... nameVariable has an average rating, and it is a Chinese place and expensive near nearVariable.}.

Here is some question: {nameVariable cheap excellent familyFriendly nearVariable}. Here is a rewrite of the question which is {UNCONSCIENTIOUSNESS} : {Yeah, I am not sure. nameVariable is cheap and kid friendly near nearVariable with a damn excellent rating!}.

Here is some question: {nameVariable pub Indian 20-25 nearVariable}. Here is a rewrite of the question which is {UNCONSCIENTIOUSNESS} : {Oh God oh gosh I don't know. nameVariable is an Indian place, nameVariable is a pub, and it is kind of near nearVariable, also it has a price range of £20-25.}.

Here is some question: {nameVariable French moderate}. Here is a rewrite of the question which is {UNCONSCIENTIOUSNESS} : {Oh gosh I mean, I am not sure. nameVariable is damn moderately priced and a French place.}.

Here is some question: {What does Apple do?}. Here is a rewrite of the question which is {UNCONSCIENTIOUSNESS} : {

**Model Output**

Oh gosh, erm... you know... what is it that Apple does again?